\documentclass[aps,preprint,showpacs,amsmath,amssymb]{revtex4-1}
\usepackage[dvips]{graphicx}
\usepackage{textcomp}
\usepackage{natbib}
\begin{document}
\newcommand {\nc} {\newcommand}
\nc {\beq} {\begin{eqnarray}}
\nc {\eol} {\nonumber \\}
\nc {\eeq} {\end{eqnarray}}
\nc {\eeqn} [1] {\label{#1} \end{eqnarray}}
\nc {\eoln} [1] {\label{#1} \\}
\nc {\ve} [1] {\mbox{\boldmath $#1$}}
\nc {\rref} [1] {(\ref{#1})}
\nc {\Eq} [1] {Eq.~(\ref{#1})}
\nc {\re} [1] {Ref.~\cite{#1}}
\nc {\dem} {\mbox{$\frac{1}{2}$}}
\nc {\arrow} [2] {\mbox{$\mathop{\rightarrow}\limits_{#1 \rightarrow #2}$}}

\title[A unique decay process: $\beta$ delayed emission of a proton and a neutron 
by the $^{11}$Li halo nucleus]
{A unique decay process: $\beta$ delayed emission of a proton and a neutron 
by the $^{11}$Li halo nucleus}
\author{D. Baye}  
\email{dbaye@ulb.ac.be}
\affiliation{Physique Quantique, C.P. 165/82, and 
Physique Nucl\'eaire Th\'eorique et Physique Math\'ematique, C.P. 229, \\
Universit\'e Libre de Bruxelles (ULB), B 1050 Brussels, Belgium}
\author{P. Descouvemont}  
\email{pdesc@ulb.ac.be}
\affiliation{Physique Quantique, C.P. 165/82, and 
Physique Nucl\'eaire Th\'eorique et Physique Math\'ematique, C.P. 229, \\
Universit\'e Libre de Bruxelles (ULB), B 1050 Brussels, Belgium}
\author{E. M. Tursunov}
\affiliation{Institute of Nuclear Physics, Uzbekistan Academy of Sciences, \\ 
100214, Ulugbek, Tashkent, Uzbekistan}
\email{tursune@inp.uz}
\date{\today}
\begin{abstract}
The neutron-rich $^{11}$Li halo nucleus is unique among nuclei with known separation energies 
by its ability  to emit a proton and a neutron in a $\beta$ decay process. 
The branching ratio towards this rare decay mode is evaluated within a three-body model 
for the initial bound state and with Coulomb three-body final scattering states. 
The branching ratio should be comprised between two extreme cases, 
i.e.\ a lower bound $6 \times 10^{-12}$ obtained with a pure Coulomb wave 
and an upper bound $5 \times 10^{-10}$ obtained with a plane wave. 
A simple model with modified Coulomb waves provides plausible values 
between between $0.8 \times 10^{-10}$ and $2.2 \times 10^{-10}$ 
with most probable total energies of the proton and neutron between 0.15 and 0.3 MeV. 
\end{abstract}
\maketitle
\section{Introduction}
Among their remarkable properties, nuclei with a neutron halo display unusual $\beta$-decay channels. 
There is indeed the possibility that the decay occurs in the halo, releasing the halo nucleons. 
This process has been observed in the $\beta$ delayed deuteron decay of $^6$He and $^{11}$Li 
\cite{RBG90,BJJ93,ABB02,MBG96,BGG97,RAB08}. 
It is however severely limited by the energy conservation condition 
\beq
S_{2n} < B(^2{\rm H}) + (m_n - m_p -m_e) c^2 \approx 3.007 {\rm\ MeV}
\eeqn{1.1}
where $S_{2n}$ is the two-neutron separation energy of the halo nucleus, 
$B(^2\rm{H})$ is the binding energy of the deuteron, 
and $m_n$, $m_p$, and $m_e$ are the neutron, proton, and electron masses, respectively. 
Only a few nuclei have low enough separation energies to allow this decay: 
$^6$He, $^8$He, $^{11}$Li, $^{14}$Be, $^{17}$B, $^{27}$F, \dots 

Another $\beta$ decay process is even more severely restricted, 
i.e. a decay of a halo neutron releasing a free neutron and a free proton. 
The condition is 
\beq
S_{2n} < (m_n - m_p -m_e) c^2 \approx 0.782 {\rm\ MeV}.
\eeqn{1.2}
Among nuclei with known two-neutron separation energy, the unique nucleus where 
this decay is allowed is $^{11}$Li, 
\beq
^{11}{\rm Li} \rightarrow\ ^9{\rm Li} + n + p + e^- + \tilde{\nu}_e
\eeqn{1.3}
with the separation energy \cite{GBA09} 
\beq
S_{2n} = 378 \pm 5 {\rm\ keV}.
\eeqn{1.4}
This process should be observable if the branching ratio is large enough. 
However, the small energy available for the decay indicates that the phase space 
is much smaller than for the deuteron emission. 
How rare is this decay is the main question raised in the present exploratory study. 

The $^{11}$Li nucleus is described in a $^9$Li+$n$+$n$ three-body model \cite{DDB03} 
as in our previous studies of the deuteron delayed emission \cite{BTD06,TBD06}. 
The $^9$Li+$n$+$p$ final state is in the three-body continuum of $^{11}$Be. 
The calculation of wave functions in this continuum is much more complicated 
than in the three-body continuum of $^6$He \cite{DTV98,DTB06}. 
The construction of three-body scattering states for $^9$Li+$n$+$n$ would already be more difficult 
than for $\alpha$+$n$+$n$ because of the poor knowledge of the $^9$Li+$n$ interaction. 
The study of the $^9$Li+$n$+$p$ continuum is worse for several reasons. 
(i) The halo nucleons are not identical and the wave functions have about twice as much components 
at the same level of truncation in an expansion in hyperspherical harmonics. 
(ii) The presence of a Coulomb interaction between the $^9$Li core and the proton 
requires a more complicated treatment than in the neutral case. 
(iii) The structure of the continuum wave functions is more complicated since 
one can expect a larger number of bound states to which they must be orthogonal.  
For these reasons, the technique that we have developed \cite{DTB06} can not provide 
a converged calculation with our present computer capabilities. 
Since an evaluation of the branching ratio would be necessary to guide future experiments, 
we shall simplify the study by describing the continuum with three-body Coulomb waves. 
This approximation should be accurate enough to estimate the order of magnitude 
of the branching ratio and the shape of the energy distribution. 

In section \ref{sec:theory}, we present general formulas for the decay probability 
per time unit for the $\beta$ delayed $np$ emission. 
In section \ref{sec:results}, we evaluate the branching ratio and discuss its origin. 
Concluding remarks are presented in section \ref{sec:conclusion}.
\section{Decay probability for $\beta$ delayed $np$ emission}
\label{sec:theory}
\subsection{General expression of decay probability}
In this section, we establish the general expression for the $\beta$ decay probability 
distribution for a three-body final state in the continuum. 
The initial nucleus with mass number $A$ is described as a three-body bound state of a core and two nucleons. 
This state with angular momentum $J_i$, projection $M_i$, and parity $\pi_i$ 
is expressed in hyperspherical coordinates. 
The spin, isospin and parity of the core are neglected. 
Three-body scattering states are discussed in Ref.~\cite{DTB06} and used in Ref.~\cite{BCD09}. 

Let us follow the notation in Ref.~\cite{BCD09} and denote the three particles as 1, 2, and $c$ (for the core). 
Let $\ve{k}_{12}$ be the relative wave vector between particles 1 and 2 
and $\ve{k}_{c(12)}$ be the relative wave vector between the center of mass of those particles and the core. 
When the spin of the core $c$ is neglected, 
the outgoing scattering states can be denoted as $\Psi^{(+)}_{\ve{k}_{12} \ve{k}_{c(12)}M_1 M_2}$, 
where $M_1$ and $M_2$ are the projections of the spins of particles 1 and 2. 
These states are assumed to be normalized with respect to 
$\delta (\ve{k}_{12}-\ve{k}'_{12}) \delta (\ve{k}_{c(12)}-\ve{k}'_{c(12)}) \delta_{M_1M'_1} \delta_{M_2M'_2}$. 

The distribution of decay probability per time unit can be written as 
\beq
\frac{dW}{d\ve{k}_{12}d\ve{k}_{c(12)}} = \frac{1}{2\pi^3} \frac{m_e c^2}{\hbar} G_{\beta}^2 f(Q-E) 
\frac{1}{2J_i+1} \sum_{M_i} \sum_{M_1M_2} \left( |M_{\rm F}|^2 + \lambda^2 \sum_\mu |M_{{\rm GT} \mu}|^2 \right),
\eeqn{2.1}
where $G_{\beta} \approx 2.996 \times 10^{-12}$ is the dimensionless $\beta$-decay constant, 
$\lambda \approx -1.268$ is the ratio of the axial-vector to vector coupling constants, 
and $E$ is the total energy of the nuclear fragments. 
The Fermi integral $f(Q-E)$ depends on the kinetic energy $Q-E$ available for the electron and antineutrino 
with 
\beq
Q = (m_n - m_p -m_e) c^2 - S_{2n}.
\eeqn{2.2}
The Fermi and Gamow-Teller matrix elements are respectively given by 
\beq
M_{\rm F}(E) = \langle \Psi^{(-)}_{\ve{k}_{12} \ve{k}_{c(12)}M_1 M_2}|\sum_{j=1}^2 t_{j-} | \Psi^{J_iM_i\pi_i} \rangle
\eeqn{2.3}
and
\beq
M_{{\rm GT}\mu}(E) = 2\langle \Psi^{(-)}_{\ve{k}_{12} \ve{k}_{c(12)}M_1 M_2}|\sum_{j=1}^2 t_{j-} s_{j\mu}| \Psi^{J_iM_i\pi_i} \rangle
\eeqn{2.4}
where $\ve{s}_j$ and $\ve{t}_j$ are the spin and isospin of particle $j$, 
and $\mu=-1$, 0, +1 labels the tensor components of the spin. 

If one integrates expression \rref{2.1} over all directions, 
the distribution of probability as a function of the total center-of-mass energy $E < Q$ 
of the three particles is given by 
\beq
\frac{dW}{dE} = \frac{1}{2\pi^3} \frac{m_e c^2}{\hbar} G_{\beta}^2 f(Q-E) 
\left( \frac{dB(\rm F)}{dE} + \lambda^2 \frac{dB(\rm GT)}{dE} \right).
\eeqn{2.5}
The Fermi and Gamow-Teller strengths appearing in this expression are given for $\sigma$ = F or GT by 
\beq
\frac{dB(\sigma)}{dE} = \frac{1}{2J_i+1} \int d\ve{k}_{12} \int d\ve{k}_{c(12)} 
\delta \left(E - \frac{\hbar^2 k_{12}^2}{2\mu_{12}} - \frac{\hbar^2 k_{c(12)}^2}{2\mu_{c(12)}} \right) 
\sum_{M_i} \sum_{M_1M_2} \sum_\mu |M_{\sigma\mu}|^2 
\eeqn{2.6}
where $\mu_{12}$ is the reduced mass of particles 1 and 2, 
and $\mu_{c(12)}$ is the reduced mass of the core $c$ and the system 1+2.

The total transition probability per time unit $W$ is obtained by integrating \rref{2.5} from zero to $Q$. 
The branching ratio can than be derived as 
\beq
{\cal R} = W t_{1/2}/\ln 2,
\eeqn{2.7}
where $t_{1/2} \approx 8.75$ ms is the half life of $^{11}$Li. 
\subsection{Bound-state and scattering three-body wave functions}
In hyperspherical coordinates, 
the three-body wave function of a bound state is defined as 
\beq
\Psi^{JM\pi}(\rho,\Omega_{5\rho}) = \rho^{-5/2} \sum_{\gamma K} {\chi}^{J\pi}_{\gamma K}(\rho)\ 
{\cal Y}^{JM}_{\gamma K}(\Omega_{5\rho}),
\eeqn{3.0}
where $\rho$ is the hyperradius, $\Omega_{5\rho}$ represents the five hyperangles, 
and ${\cal Y}^{JM}_{\gamma K}(\Omega_{5\rho})$ is a hyperspherical harmonics 
(see \re{DDB03} for definitions and notations). 
The symbol $K$ corresponds to the hypermomentum quantum number 
and $\gamma$ is a shorthand notation for $l_x l_y L S$, 
where $L$ is the total orbital momentum, $S$ is the total spin, 
and $l_x$ and $l_y$ are the orbital momenta for the relative motions 
corresponding to the Jacobi coordinates between particles 1 and 2 and between 
the core and the center of mass of 1+2, respectively. 
The parity of these relative motions is given by $\pi = (-1)^{l_x+l_y} = (-1)^K$, 
which implies that $K$ must be either even or odd. 
The hyperradial wave functions are expanded as 
\beq
{\chi}^{J\pi}_{\gamma K}(\rho) = \sum_{i=1}^N c^{J\pi}_{\gamma Ki} f_i(\rho)
\eeqn{3.1}
in terms of $N$ Lagrange functions $f_i$ (see \re{DDB03} for definitions). 
Since the hyperspherical harmonics and the Lagrange functions are orthonormal, 
the wave functions are normed if 
\beq
\sum_{\gamma K} \sum_{i=1}^N \left( c^{J\pi}_{\gamma Ki} \right)^2 = 1.
\eeqn{3.2}
In the present approximation of $^{11}$Li with a $0^+$ core, 
particles 1 and 2 are neutrons. 
The effective angular momentum and parity are $J^\pi = 0^+$. 
The isospin is $T = 1$ for the halo neutrons. 
Since they are identical, antisymmetry imposes $(-1)^{l_x} = (-1)^S$. 

The final states of the decay are three-body scattering states. 
It is convenient to replace the projections $M_1$ and $M_2$ 
by the total spin $S$ of nucleons 1 and 2 and its projection $\nu$. 
With a zero-spin core, $S$ is the channel spin. 
The ingoing scattering states read \cite{BCD09} 
\beq 
\Psi^{(-)}_{\ve{k_{12}} \ve{k_{c(12)}} S\nu} 
= (2\pi)^{-3} \rho^{-5/2} \left( \frac{A}{A_c} \right)^{3/4} \sum_{JM} 
\sum_{l_{x\omega} l_{y\omega} L_\omega K_\omega} (L_\omega S\, M\!-\!\nu\, \nu | J M) 
{\cal Y}_{l_{x\omega} l_{y\omega} K_\omega}^{L_\omega M-\nu\,*} (\Omega_{5k}) 
\eol \times
\sum_{\gamma K} (-1)^K {\cal Y}_{\gamma K}^{JM}(\Omega_{5\rho}) 
\chi^{J\pi *}_{\gamma K (\gamma_\omega K_\omega)} (\rho),
\eeqn{3.3}
where $A_c = A-2$ is the core mass number. 
This formula differs from \re{BCD09} because of a different normalization. 
The normalization for the hyperradial partial waves is \cite{DTB06} 
\beq
{\chi}^{J\pi}_{\gamma K (\gamma_\omega K_\omega)}(\rho) 
\mathop{\rightarrow} \limits_{\rho \rightarrow \infty} 
i^{K_\omega+1}(2\pi/k)^{5/2} 
\left[ H^-_{\gamma K+2} (k\rho) \delta_{\gamma \gamma_\omega}\delta_{KK_\omega}
-U^{J\pi}_{\gamma K,\gamma_\omega K_\omega} H^+_{\gamma K+2} (k\rho) \right].
\eeqn{3.4}
In this expression, the wave number $k$ is given by $\sqrt{2m_N E/\hbar^2}$, 
where $m_N$ is the nucleon mass, 
and $U^{J\pi}_{\gamma K,\gamma_\omega K_\omega}$ is an element of 
the infinite-dimensional collision matrix. 
The subscript $\omega$ refers to the entrance channel. 
Let us recall here that, in a three-body scattering state, there is in principle 
an infinity of degenerate entrance channels. 

For charged systems, one has 
\beq
H^{\pm}_{\gamma K+2}(x)&=&G_{K+\frac{3}{2}}(\eta_{\gamma K},x) \pm iF_{K+\frac{3}{2}}(\eta_{\gamma K},x),
\label{3.5}
\eeq
where $G_{K+3/2}$ and $F_{K+3/2}$ are the irregular and regular Coulomb functions, respectively \cite{TB86}. 
The Sommerfeld parameters $\eta_{\gamma K}$ are given by 
\beq
\eta_{\gamma K} = Z^{J\pi}_{\gamma K,\gamma K} \frac{m_N e^2}{\hbar^2 k},
\label{3.6}
\eeq
where $Z^{J\pi}_{\gamma K,\gamma K}$ is a diagonal element of the effective-charge matrix 
and depends thus on the channel. 
One usually neglects non-diagonal terms of this matrix \cite{VNA01}. 

In the neutral case $\eta_{\gamma K} = 0$, \Eq{3.5} reduces to an expression independent of $\gamma$, 
\beq
H^{\pm}_{\gamma K+2} (x)=\pm i \left( \frac{\pi x}{2} \right) ^{1/2}
\left[ J_{K+2}(x)\pm i Y_{K+2}(x) \right],
\label{3.7}
\eeq
where $J_n (x)$ and $Y_n (x)$ are Bessel functions of first and second kind, respectively.
\subsection{Reduced transition probabilities}
For $^{11}$Li, with the spin of the core neglected, we assume $J_i = M_i = 0$. 
The final state is a three-body $^9$Li+$n$+$p$ scattering state \rref{3.3}. 
Selection rules restrict this state to its $0^+$ and $1^+$ components 
for the Fermi and Gamow-Teller transitions, respectively. 
In the present approximation, the properties of the final state only depend 
on the total spin $S$ and isospin $T$ of the two nucleons. 
For the nucleons in the $^9$Li+$n$+$p$ continuum of $^{11}$Be, 
the isospin is given by $(-1)^{l_x + S + T} = -1$. 
For $S=0$, $l_x$ even corresponds to $T=1$ and $l_x$ odd to $T=0$. 
For $S=1$, $l_x$ even corresponds to $T=0$ and $l_x$ odd to $T=1$. 
The number of channels is thus about the double of the number 
of channels in the $^9$Li+$n$+$n$ continuum of $^{11}$Li.

The sum over $M_1$ and $M_2$ in \Eq{2.4} can be replaced by a sum over the channel spin 
equal to $S$ and its projection $\nu$. 
If one replaces the wave vectors $\ve{k}_{12}$ and $\ve{k}_{c(12)}$ by their hyperspherical 
counterparts $k$ and $\Omega_{5k}$ \cite{BCD09}, the reduced transition probabilities can be written as 
\beq
\frac{dB(\sigma)}{dE} = \dem E^2 \left( \frac{2m_N}{\hbar^2} \right)^3 
\sum_{S\nu\mu} \int d\Omega_{5k} |M_{\sigma\mu}|^2, 
\eeqn{4.1}
where $\mu = 0$ for F and $\mu = -1$, 0, 1 for GT.

After integration over $\Omega_{5\rho}$, the matrix elements can be written as 
\beq
M_{\sigma\mu} 
= \sqrt{2} (2\pi)^{-3} \sum_{l_{x_\omega} l_{y_\omega} L_\omega K_\omega} (L_\omega S\,\mu\!-\!\nu\, \nu|J \mu) 
{\cal Y}^{L_\omega \mu-\nu *}_{l_{x_\omega} l_{y_\omega} K_\omega} (\Omega_{5k}) 
I^{J \pi}_{l_{x_\omega} l_{y_\omega} L_\omega S K_\omega}(\sigma),
\eeqn{4.2}
where the spherical harmonics depend on the hyperangles characterizing the wave vectors, 
i.e.\ they depend on the directions of emission of the core and nucleons, 
and on the repartition of the total energy $E$ between these particles \cite{BCD09}. 
The expressions $I^{J \pi}_{l_{x_\omega} l_{y_\omega} L_\omega S K_\omega}(\sigma)$ 
are one-dimensional integrals over the hyperradius $\rho$. 
After integration over $\Omega_{5k}$ and summation over the projections $\mu$ and $\nu$, 
the reduced transition probabilities simplify as 
\beq
\frac{dB(\sigma)}{dE} = \frac{2J+1}{(2\pi)^6} E^2 \left( \frac{2m_N}{\hbar^2} \right)^3 
\sum_{l_{x_\omega} l_{y_\omega} L_\omega S K_\omega} 
\left| I^{J \pi}_{l_{x_\omega} l_{y_\omega} L_\omega S K_\omega}(\sigma) \right|^2.
\eeqn{4.3}

Let us list the possible cases. 
For the Fermi operator, the scattering-state partial wave has $J=0$ and $\pi=+1$. 
One obtains for $S = 0$, 
\beq
I^{0^+}_{l_{x_\omega} l_{y_\omega} 0 0 K_\omega}({\rm F}) 
= \sum_{l_x {\rm\ even}} \sum_K 
\int_0^\infty \chi^{0^+}_{l_xl_x00K(l_{x_\omega} l_{y_\omega} 00 K_\omega)} (\rho) 
\chi^{0^+}_{l_xl_x00K} (\rho) d\rho,
\eeqn{4.4}
and for $S = 1$, 
\beq
I^{0^+}_{l_{x_\omega} l_{y_\omega} 1 1 K_\omega}({\rm F}) 
= \sum_{l_x {\rm\ odd}} \sum_K 
\int_0^\infty \chi^{0^+}_{l_xl_x11K(l_{x_\omega} l_{y_\omega} 11 K_\omega)} (\rho) 
\chi^{0^+}_{l_xl_x11K} (\rho) d\rho.
\eeqn{4.5}
For the Gamow-Teller operator, the scattering-state partial wave has $J=1$ and $\pi=+1$.
One obtains for $S = 0$, 
\beq
I^{1^+}_{l_{x_\omega} l_{y_\omega} 1 0 K_\omega}({\rm GT}) 
= \sqrt{\frac{1}{3}} \sum_{l_x {\rm\ odd}} \sum_K 
\int_0^\infty \chi^{1^+}_{l_xl_x10K(l_{x_\omega} l_{y_\omega} 10 K_\omega)} (\rho) 
\chi^{0^+}_{l_xl_x11K} (\rho) d\rho,
\eeqn{4.6}
and for $S = 1$, 
\beq
I^{1^+}_{l_{x_\omega} l_{y_\omega} L_\omega 1 K_\omega}({\rm GT}) 
= -\sum_{l_x {\rm\ even}} \sum_K 
\int_0^\infty \chi^{1^+}_{l_xl_x01K(l_{x_\omega} l_{y_\omega} L_\omega 1 K_\omega)} (\rho) 
\chi^{0^+}_{l_xl_x00K} (\rho) d\rho
\eol
- \sqrt{\frac{2}{3}} \sum_{l_x {\rm\ odd}} \sum_K 
\int_0^\infty \chi^{1^+}_{l_xl_x11K(l_{x_\omega} l_{y_\omega} L_\omega 1 K_\omega)} (\rho) 
\chi^{0^+}_{l_xl_x11K} (\rho) d\rho.
\eeqn{4.7}
Because of the properties of Lagrange functions, the integrals are simply given by 
\beq
\int_0^\infty \chi^{J^\pi}_{\gamma K(\gamma_\omega K_\omega)} (\rho) \chi^{0^+}_{\gamma K} (\rho) d\rho 
\approx \sum_i (h\lambda_i)^{1/2} c^{0^+}_{\gamma Ki} \chi^{J\pi}_{\gamma K(\gamma_\omega K_\omega)} (hx_i), 
\eeqn{4.8}
where $x_i$ and $\lambda_i$ are the zeros and weights of the Gauss quadrature 
associated with the Lagrange functions and $h$ is a scaling factor 
providing mesh points $\rho_i = h x_i$ adapted to the extension of the physical system. 
\subsection{Coulomb-wave approximation}
As mentioned in the introduction, we shall use a simpler approximation based on 
three-body Coulomb functions. 
In the pure Coulomb case, the scattering partial waves are approximated as 
\beq
{\chi}^{J\pi}_{\gamma K (\gamma_\omega K_\omega)}(\rho) 
= 2 i^K (2\pi/k)^{5/2} F_{K+3/2} (\eta_{\gamma K}, k\rho) \delta_{\gamma \gamma_\omega}\delta_{KK_\omega}.
\eeqn{5.2}
With this approximation, the reduced transitions probabilities become 
\beq
\frac{dB({\rm F})}{dE} = \frac{4m_N}{\pi k \hbar^2} 
\sum_{l_{x_\omega}} \sum_{K_\omega} \left| \int_0^\infty F_{K_\omega+3/2} (\eta_{\gamma_\omega K_\omega}, k\rho) 
\chi^{0^+}_{\gamma_\omega K_\omega} (\rho) d\rho  \right|^2 
\eeqn{5.3}
where $\gamma_\omega$ represents here $l_{x_\omega}l_{x_\omega}SS$ with $(-1)^S = (-1)^{l_{x_\omega}}$, and 
\beq
\frac{dB({\rm GT})}{dE} = 3 \frac{dB({\rm F})}{dE}.
\eeqn{5.4}
The F and GT reduced transition probabilities are then proportional. 
\section{Results and discussion}
\label{sec:results}
\subsection{$Q$ value and Fermi integral}
With the separation energy \rref{1.4} of $^{11}$Li, the $Q$ value 
for the $\beta$ delayed $np$ emission is quite small, 
\beq
Q \approx 0.404 \rm{\ MeV}.
\eeqn{6.1}
Moreover, the wave number is also small, 
\beq
k < 0.14  \rm{\ fm}^{-1}.
\eeqn{6.2}
This will affect the behavior of wave functions at small distances.

\begin{figure}[hbt]
\setlength{\unitlength}{1mm}
\begin{picture}(140,80) (0,0) 
\put(-10,-60){\mbox{\scalebox{0.5}{\includegraphics{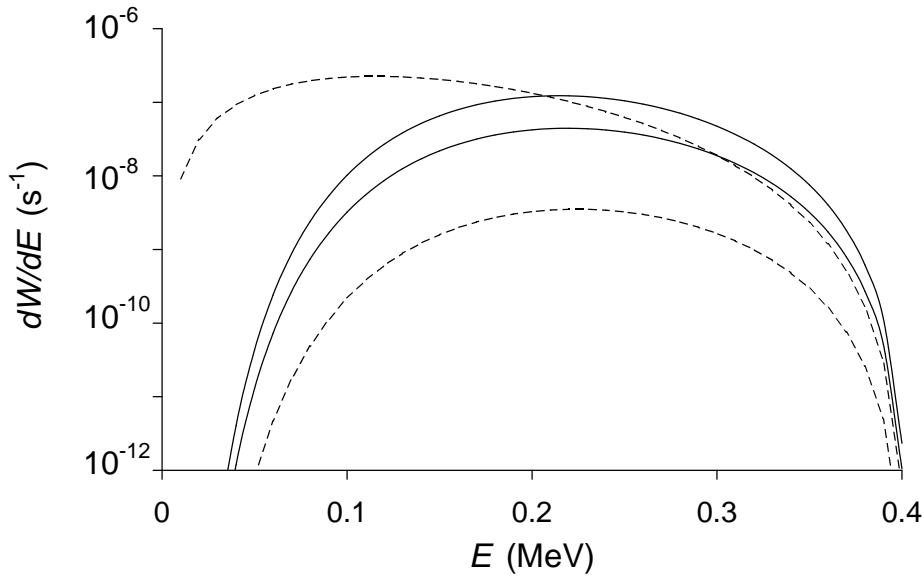}}}}
\end{picture} \\
\caption{Fermi integral $f(Q-E)$ as a function of the total energy $E$ 
of the emitted nuclear fragments for the hydrogen isotopes $^1$H, $^2$H, and $^3$H.}
\label{fig:1}
\end{figure}

In Fig.~\ref{fig:1}, the Fermi integrals $f(Q-E)$ for the emission of the 
different hydrogen isotopes are compared. 
The emitted electron being much faster than the heavy particles, 
the charge $Z=4$ is used in the electron attraction by the final nuclear system. 
The $Q$ values are 2.63 and 4.82 MeV for $^2$H and $^3$H, respectively. 
Both processes have been observed experimentally. 
In spite of a much larger Fermi integral, the branching ratio for tritons \cite{LDE84,MBA09}
is not larger than for deuterons \cite{MBG96,BGG97}. 
The emission of deuterons can be fairly well described in a model 
where the $^9$Li+$d$ resonance observed in the model of \re{TBD06} 
is shifted to about 0.8 MeV and an absorption towards other open channels is included \cite{TBD10}. 
To our knowledge, no model description of the $\beta$ delayed triton emission is available. 
The difficulty comes from the fact that this decay can not be described in a three-body model. 

The Fermi integral for $\beta$ delayed proton-neutron decay is much smaller than the other ones 
because of the limited phase space. 
The branching ratio can thus be expected to be much smaller than for the other $\beta$ delayed emissions. 
\subsection{Bound-state and Coulomb wave functions}
The $^{11}$Li ground state is obtained as a bound state in a $^9$Li+$n$+$n$ model. 
The Minnesota force is used as nucleon-nucleon interaction \cite{TLT77}. 
The $^9$Li-neutron interaction is the P2 interaction of \re{TZ94}, 
downscaled by a factor 0.97 to approximate the experimental binding energy. 
The $s$ and $p3/2$ forbidden states are eliminated by supersymmetric transformations 
\cite{Ba87}. 
The sum over partial waves in the wave function \rref{3.0} is restricted 
to $K \le K_{\rm max} = 20$. 
This wave function involves 66 components. 
The hyperradial functions \rref{3.1} are expanded over a Lagrange-Laguerre basis 
with integrals calculated with the corresponding Gauss-Laguerre quadrature, 
as explained in \re{DDB03}. 
The calculation is performed with $N = 40$ basis functions and mesh points and 
the mesh is scaled with a factor $h = 0.4$ (see \re{DDB03} for definitions). 
The resulting energy is $-0.391$ MeV, close to the experimental value. 
We use the experimental $Q$ value \rref{6.1} in the calculation of Fermi integrals. 

Because of the low values \rref{6.2} of the wave number $k$, 
the hyperradial scattering wave functions should be small at distances 
where the integrals \rref{4.5}-\rref{4.8} are significant. 
They become smaller and smaller with increasing hypermomentum $K$. 
For this reason, the sum in Eqs.~\rref{4.3} or \rref{5.3} is strongly dominated by $K = 0$. 
The $K = 0$ component of the ground state thus plays a crucial role. 
One should however not expect to use a low value of $K_{\rm max}$ 
because the convergence of this $K = 0$ component is slow \cite{DDB03}. 

For the three-body scattering states, we use approximations based on Coulomb waves. 
Let us first evaluate the effective charges entering the Sommerfeld parameter. 
Because of the $K = 0$ dominance in expression \rref{5.3}, 
we can restrict ourselves to this value 
and average the Coulomb potential over the $K=0$ hyperspherical harmonics. 
The Coulomb potential is simply 
\beq
V_C = \frac{3e^2}{|\ve{r}_c - \ve{r}_1|},
\eeqn{6.3}
where subscript 1 corresponds here to the proton. 
Using the hyperradius $\rho$ and the hyperangle $\alpha$ \cite{DDB03}, 
the $K=0$ average can be written as 
\beq
\frac{Z_{00,00}^{0^+} e^2}{\rho} & = & \sqrt{\mu_{c1} } 
\left\langle Y^{00}_{00} \bigg| \frac{3e^2}{\rho \cos \alpha} \bigg| Y^{00}_{00} \right \rangle 
\eol
& = & \sqrt{\mu_{c1} } \frac{3e^2}{\rho}\, \frac{16}{\pi} \int_0^{\pi/2} \sin^2 \alpha \cos \alpha d\alpha, 
\eeqn{6.4}
where $\mu_{c1}=A_c/(A_c+1)$ is the reduced mass of the core and the proton 
and $\gamma=0$ represents $l_x = l_y = L = S = 0$.  
Hence, the effective charge reads
\beq
Z_{00,00}^{0^+} = \frac{48}{\pi \sqrt{10}} \approx 4.83.
\eeqn{6.5}
To simplify a calculation dominated by $K=0$, we shall use this effective value for all partial waves. 
\subsection{Distribution of decay probability per time unit}
\begin{figure}[ht]
\setlength{\unitlength}{1mm}
\begin{picture}(140,85) (0,0) 
\put(  0,-50){\mbox{\scalebox{0.5}{\includegraphics{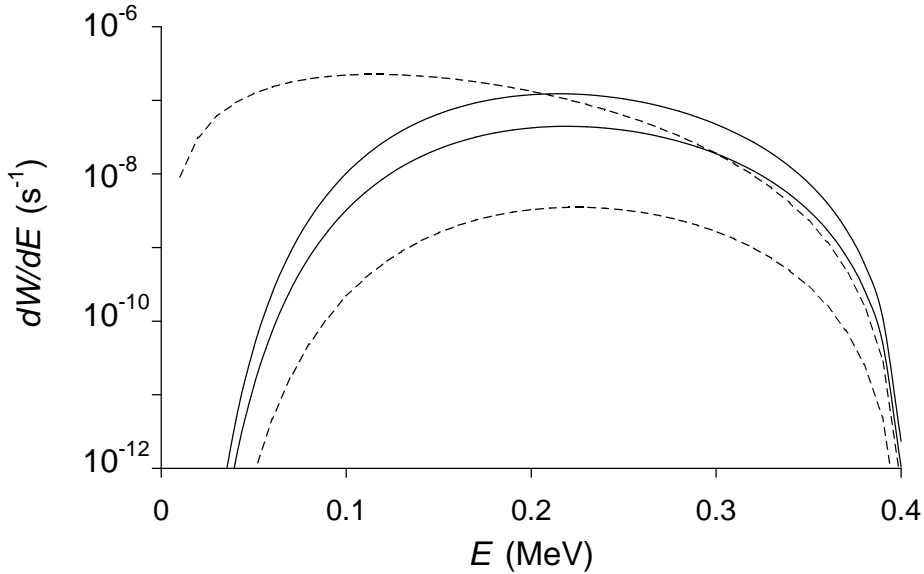}}}}
\end{picture} \\
\caption{Distribution of decay probability per time unit 
for the $\beta$ delayed $np$ decay of $^{11}$Li: 
plane wave (upper dashed curve), Coulomb wave with effective charge \rref{6.5} (lower dashed curve), 
and shifted Coulomb waves with $a = 10$ (lower full curve) and 15 fm (upper full curve).}
\label{fig:2}
\end{figure}
Various approximations of the distribution of decay probability per time unit 
for the $\beta$ delayed $np$ decay of $^{11}$Li are displayed in Fig.~\ref{fig:2}. 
With the effective charge \rref{6.5}, one obtains the lower dashed curve 
giving the total probability $W = 5.1 \times 10^{-10}$ s$^{-1}$ 
and thus the branching ratio ${\cal R} = 6.5 \times 10^{-12}$. 
These results can be contrasted with a plane-wave calculation ($\eta_{00} = 0$) 
which leads to the upper dashed curve giving $W = 3.8 \times 10^{-8}$ s$^{-1}$ 
and ${\cal R} = 4.8 \times 10^{-10}$. 
The Coulomb-wave calculation is pessimistic because it neglects 
an enhanced probability of presence of the emitted nucleons at short distances 
due to the attractive nuclear interaction. 
The plane-wave calculation overestimates the probability of presence of the emitted 
proton at short distances because of the missing Coulomb repulsion by the nucleus. 
Both calculations neglect a possible absorption towards other open channels 
affecting the final wave function. 
However, it is difficult to figure out whether one of these cases 
is a better approximation. 
Hence we turn to a slightly different approach. 

\begin{figure}[hbt]
\setlength{\unitlength}{1mm}
\begin{picture}(140,75) (0,0) 
\put(  0,-60){\mbox{\scalebox{0.5}{\includegraphics{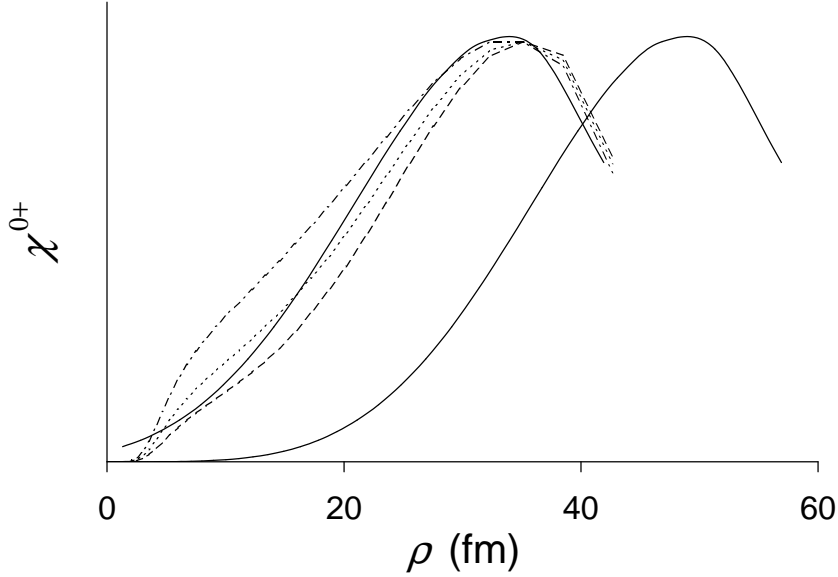}}}}
\end{picture} \\
\caption{$K = 0$ components of the lowest positive-energy pseudostate near 0.4 MeV 
for $K_{\rm max} = 12$ (dashed line), 16 (dotted line), and 20 (dash-dotted line) 
normalized to the $K = 0$ Coulomb wave with $Z_{00,00}^{0^+} = 4.83$ (right full line); 
same Coulomb wave shifted by 15 fm (left full line). }
\label{fig:3}
\end{figure}
For a better but still simple approximation based on Coulomb functions, 
we have considered the eigenstates of the $^9$Li+$n$+$p$ system. 
With $K_{\rm max} = 20$, its wave functions involve 121 components. 
The conditions of the calculation are the same as for $^9$Li+$n$+$n$ 
except for the additional Coulomb interaction \rref{6.3} between $^9$Li and $p$ 
and a reduced symmetry. 
The $^9$Li+$p$ relative motion only requires the elimination of an $s$ forbidden state. 

We obtain four bound states at $-12.027$, $-3.944$, $-0.876$, and $-0.786$ MeV 
with respect to the $^9$Li+$n$+$p$ threshold. 
Even the lowest bound state is far above the experimental ground-state energy $-20.14$ MeV. 
The state at $-0.876$ is the isobaric analog of the $^{11}$Li ground state. 
The lowest positive-energy state is located at 0.379 MeV. 
It must not be considered as a resonance but rather as a pseudostate, 
a bound-state approximation of a scattering state at this energy. 
Its wave function will be useful to construct a better exploratory approximation. 

The $K = 0$ components of the lowest positive-energy state located near 0.4 MeV 
obtained with $K_{\rm max} = 12$, 16, and 20 are displayed in Fig.~\ref{fig:3}. 
The energies do not vary much with $K_{\rm max}$ but the wave function is not yet converged. 
The amplitudes of the three curves are normalized to the $K = 0$ Coulomb wave \rref{5.2} 
corresponding to the charge \rref{6.5} (right full line). 
One observes a significant shift between the pseudostate and the Coulomb wave. 
As a simple qualitative approximation, we shift the Coulomb wave by 15 fm 
towards shorter distances (left full line). 
The resulting curve simulates the general behavior of the pseudostate. 
A shift by 10 fm would also be plausible. 

We thus simulate the $K=0$ component 
of the scattering state with the shifted Coulomb functions
\beq
{\chi}^{J\pi}_{\gamma K (\gamma_\omega K_\omega)}(\rho) 
= 2 i^K (2\pi/k)^{5/2} F_{K+3/2} [\eta_{\gamma K}, k(\rho + a)] \delta_{\gamma \gamma_\omega}\delta_{KK_\omega}
\eeqn{6.6}
with $a = 10$ and 15 fm. 
These functions do not vanish at the origin but this drawback has little influence, 
i.e.\ a smaller influence than other approximations. 
The results are displayed as full curves in Fig.~\ref{fig:2}: 
the lower curve corresponds to $a = 10$ fm and 
the upper curve corresponds to $a = 15$ fm. 
Their maximum is slightly shifted towards higher energies. 
The most probable total energies $E$ are located between 0.15 and 0.3 MeV 
and the most probable total energies of the proton and neutron should approximately 
lie in the same interval since the $^9$Li core is heavier. 
This approximation corresponds to $0.6 \times 10^{-8} < W < 1.8 \times 10^{-8}$ s$^{-1}$ 
and $0.8 \times 10^{-10} < {\cal R} < 2.2 \times 10^{-10}$. 
\section{Conclusion}
\label{sec:conclusion}
In this paper, we evaluate the order of magnitude of the branching ratio for the 
$\beta$ delayed $np$ emission by $^{11}$Li, a very exotic decay process, unique among nuclei 
with known two-neutron separation energies. 
We have established the theoretical formulas for the Fermi and Gamow-Teller transitions 
leading to three-body final states. 

An accurate model calculation is made very difficult by the need of three-body scattering 
states involving three different particles, two of them charged, at very low energies 
and by our lack of knowledge of physical properties of this three-body continuum 
and of absorption effects in the final three-body channel. 
To circumvent these difficulties in an exploratory calculation, 
we have made several simplifying approximations. 
Simple models of the final state involving a plane wave and a pure Coulomb wave 
provide likely upper and lower bounds of the branching ratio, respectively. 
We think that more reasonable estimates of the branching ratio and of the energy distribution 
of the decays are obtained with shifted three-body Coulomb functions. 

The obtained branching ratio should be comprised between $6 \times 10^{-12}$ 
and $5 \times 10^{-10}$ with more plausible values 
between $0.8 \times 10^{-10}$ and $2.2 \times 10^{-10}$. 
The most probable total energies of the proton and neutron should lie between 0.15 and 0.3 MeV. 
In any case, the branching ratio is much smaller than for the deuteron 
and triton channels, i.e. $(1.3 \pm 0.13) \times 10^{-4}$ \cite{RAB08} 
and $(0.93 \pm 0.08) \times 10^{-4}$ \cite{MBA09}, respectively. 
It is even much smaller than for the hindered deuteron decay of $^6$He, 
$(2.6 \pm 1.3) \times 10^{-6}$ \cite{ABB02}. 
The main cause of this smallness is the small $Q$ value of the process 
which leads to a limited phase space. 
The observation of this $\beta$ delayed decay mode, if it is possible, 
will thus require high radioactive beam intensities and long measurement times 
to reach a significant enough number of $^{11}$Li decays. 

If this unique decay process is studied experimentally, a better model calculation 
will become necessary, with a full calculation of the three-body $^9$Li+$n$+$p$ continuum 
wave functions, using the formalism developed in Sec.~\ref{sec:theory}. 
This study should be performed with $^9$Li+$n$ and $^9$Li+$p$ optical potentials 
in order to take absorption effects into account. 
\section*{Acknowledgments}
This text presents research results of BriX (Belgian Research Initiative on eXotic nuclei), 
the interuniversity attraction pole programme P6/23 initiated by the Belgian-state 
Federal Services for Scientific, Technical and Cultural Affairs (FSTC). 
E.M.T. thanks the IAP programme for supporting his stay.
P.C.\ acknowledges the support of the F. R. S.-FNRS.
%
%
%
%
\end{document}